# Multiresonant Membrane Metasurfaces for Multifunctional Fingerprint Recognition and Real-time Biochemical Tracking

Quanlong Yang[1,*], Yapeng Dou[2], Dongyang Wang[3], Yihua Zhong[1], Fei Li[4], Jiajun He[5], Ying Zhang[4,*], Quan Xu[5], Junliang Yang[1], Ilya Shadrivov[6], Jiaguang Han[5], and Yuri Kivshar[6,*]

1School of Physics, Central South University, Changsha 410083, China

2School of Electronic Information, Central South University, Changsha 410083, China

3Optoelectronics Research Centre, University of Southampton, Southampton SO17 1BJ, UK

4Yunnan Key Laboratory of Opto-electronic Information Technology and Institute of Physics and Electronic Information, Yunnan Normal University, Kunming 650500, China

5Center for Terahertz Waves, Tianjin University, Tianjin 300072, China

6Research School of Physics, Australian National University, Canberra ACT 2601, Australia

*Corresponding author. Email: quanlong.yang@csu.edu.cn;  yingzhang27@ynnu.edu.cn;  yuri.kivshar@anu.edu.au

**Funding:**
This work was supported by the
National Natural Science Foundation of China (Grant No. 62205380)
Natural Science Foundation of Hunan Province (Grant No. 2024JJ6529).
Y.K. acknowledges support from the
Australian Research Council (Grant No DP210101292)
International Technology Center Indo-Pacific (ITC IPAC) via Army Research Office (contract FA520923C0023).

This work is under review in Advanced Materials.

**Abstract**

Label-free identification and real-time tracking of biochemical substances became critical for molecular diagnostics and chemical analysis, yet conventional resonant terahertz metasurface sensing relies on a single resonance, limiting spectral selectivity and dynamic capability. Here, we suggest multiresonant membrane metasurfaces and implement them for simultaneous static molecular fingerprint retrieval and dynamic reaction monitoring within a single pixel. We consider a membrane metasurface supporting multiple quasi-bound states in the continuum designed at target frequencies and enabling the tailoring of the field enhancement and frequency-selective interaction with target analytes. As a proof-of-concept, we achieve label-free detection of the dual fingerprint absorption features of pefloxacin at 0.78 THz and 0.99 THz, and real-time tracking of vitamin C oxidation and denaturation under ambient conditions. The kinetic profiles extracted from the THz amplitude evolution show excellent agreement with nonlinear reaction models ($R^2 > 0.92$), demonstrating quantitative biochemical tracking capabilities. Our results establish a versatile and scalable THz photonic platform that unifies static fingerprint identification and dynamic reaction monitoring, paving the way toward integrated on-chip biochemical analytics and multifunctional metasurface sensors.

# 1. Introduction

Tracking biochemical processes in real time is a fundamental goal in molecular sensing and analytical chemistry, as it offers direct insight into molecular reactions, conformational changes, and biological kinetics. Terahertz (THz) spectroscopy has emerged as a powerful label-free and non-destructive sensing approach capable of probing collective vibrational and rotational modes of biomolecules.[1-6] It was employed widely for conformation monitoring, analysis of the DNA secondary structure, and selective detection of bacterial layers based on plasmonic THz antennas.[7-10] Traditional THz sensing primarily relies on microfluidic or thin-film sample supports.[11-14] To enhance interaction with analytes, it was proposed to use chip-based waveguide platforms or employ a complementary strategy involving embedding analytes in a hydrogel matrix pre-loaded with gold nanoparticles, in order to amplify analyte responses via detecting characteristic absorption peaks.[15-16] An inherent mismatch between THz wavelengths and molecular scales leads to weak light–matter interaction posing a challenge for trace-level detection using THz spectroscopy. Recently, the use of metasurfaces allowed to improve significantly the field localization and resonance enhancement, substantially boosting the device performance.[17-24] Various metasurface-based platforms leveraging Fano resonances, localized surface plasmon polaritons, and high-Q cavity modes have been proposed to further enhance sensing performance.[25-31] However, despite these advances, achieving dynamic biochemical detection and multi-band fingerprint recognition within a single metasurface remains a major challenge.

To address this issue, we suggest a novel type of *multiband multiresonant metasurfaces* supporting quasi-bound states in the continuum (BICs). The use of quasi-BICs with finite values of the quality factor (Q factor) enables their efficient experimental excitation for high-sensitivity biochemical detection. The metasurface-based quasi-BIC resonances have garnered increasing interest in biomedical sensing by integrating the exceptional field enhancement and spectral selectivity. For instance, recent studies have demonstrated their utility in diverse applications, including lung cancer cell and cardiac troponin detection, multidimensional biochemical monitoring, mixture component analysis, and high-specificity fingerprint retrieval of α-lactose.[32-42] Despite that progress, a critical limitation of THz biosensing remains in the use of a single absorption peak shift induced by concentration variations, where the substance-specific recognition is still lacking. Conventional molecular fingerprinting relies on large-scale pixelated arrays, and obtaining a full fingerprint spectrum typically requires scanning across multiple spatial elements. [34, 43-44] Meanwhile, recent advances in THz metasurfaces have demonstrated a wide variety of high-Q resonances engineering strategies to enable broadband spectroscopy and sensitive biomolecular detection across a wide frequency range.[45-48] In contrast, the implementation of multiple resonances in a single-pixel metasurface offers a promising route for material-specific fingerprint identification through characteristic THz spectral features. Moreover, dynamic biochemical reaction tracking remains a key challenge in photonic sensing platforms, where the simultaneous capture of molecular fingerprints and reaction kinetics is rarely achieved within a single device.

Here, we report a THz trace sensing and real-time biochemical reaction monitoring platform based on engineered multiple resonances. We employ pairs of horseshoe-shaped resonators that enable the simultaneous excitation of multiple quasi-BIC modes, providing enhanced spectral tunability and strong field confinement. The metasurface is specifically designed to target Pefloxacin molecules, providing strong field enhancement at 0.78 and 0.99 THz. To validate our approach, we characterize a multiresonant membrane metasurface using THz time-domain spectroscopy, successfully observing multiple resonances that enable accurate retrieval of Pefloxacin fingerprints within a compact single-pixel device. As a demonstration, we demonstrate the capability of this platform for real-time monitoring of vitamin C oxidation and denaturation under ambient conditions, establishing its effectiveness for dynamic, label-free tracking of biochemical reactions. We believe that multifunctional metasurfaces offer an efficient solution for both molecular fingerprint identification and real-time biochemical reaction monitoring, paving the way for advanced THz sensing and wavefront engineering.

## 2. Results

### 2.1 Metasurfaces for THz sensing

**Figure 1**a illustrates the schematic of the designed terahertz sensing platform. The metasurface consists of a dual horseshoe-shaped resonator (DHR) with identical ring widths and aperture diameters, where all meta-atoms are located on a free-standing and low-loss membrane substrate of cyclic olefin copolymer (COC) of the thickness of 50 μm. The dielectric constant of COC at 1 THz is $\varepsilon = 2.32 + 0.0147i$. In the x-y plane of Figure 1c, the copper meta-atoms are arranged with the period of $P_x = P_y = 250$ μm, where the distance between the centers of the two rings is $k = 125$ μm, and the radii of SRRs are $r_1 = 36$ μm and $r_2 = 20$ μm, respectively.

We optimized the metasurfaces via introducing the orientation degree of freedom in the DHR that is defined by the angle $\varphi$ to simultaneously achieve engineered multiple resonances at the frequencies of interest (one horseshoe-shaped resonator in the unit cell is rotated clockwise, and another counter-clockwise, keeping the mirror symmetry along the *x*-direction to protect the BICs). Upon terahertz wave incidence, the metasurface supports strongly localized electromagnetic fields that amplify terahertz–matter interactions at multiple resonant frequencies of interest. This facilitates the concurrent detection of molecular vibrational fingerprints across those characteristic frequencies. To validate the efficacy of this design, pefloxacin was selected as a representative analyte for trace-level fingerprint detection. As shown in Figure 1a, pronounced fingerprint spectra are observed at specific frequencies corresponding to the characteristic vibrational modes of pefloxacin, demonstrating the ability of the metasurface to resolve multi-band molecular signatures within a single measurement.

It is worth noting that the strong light-matter interaction of resonant metasurfaces makes it highly suitable for monitoring dynamic biochemical reactions. By tracking variations in the amplitude of BIC resonances, the progression of biochemical reactions can be quantitatively assessed, offering a novel platform for label-free, real-time analysis. To demonstrate the capability of our system in capturing dynamic processes, we investigated the oxidation of vitamin C in ambient air as a representative reaction, which could be seen in Figure 1a. Unlike conventional single-mode resonances that support only one resonant peak, multiple resonances enable static fingerprint sensing and dynamic reaction tracking.

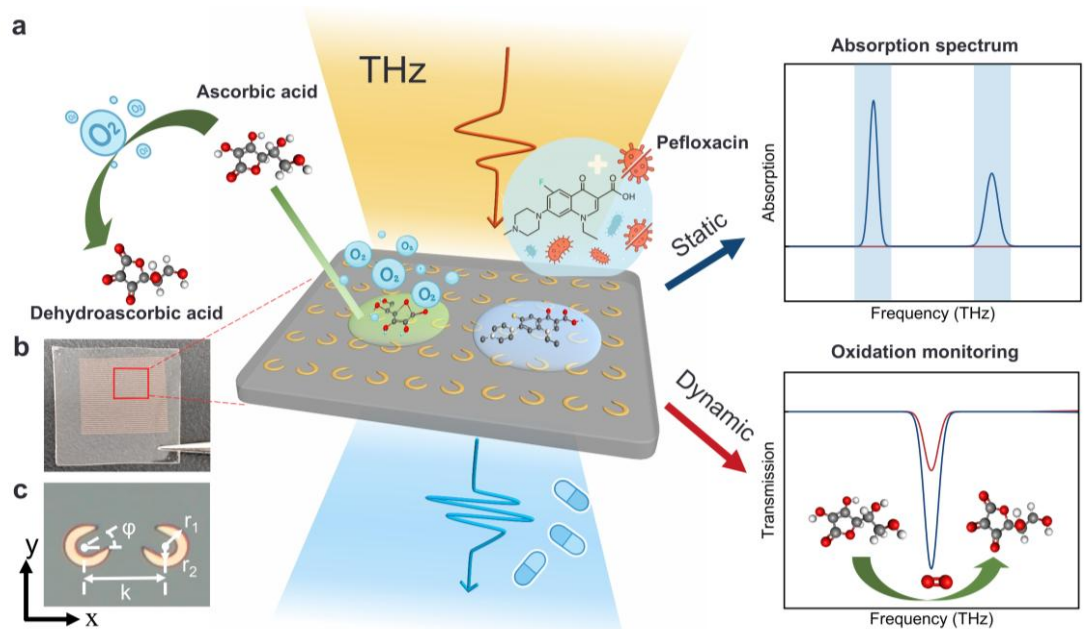

**Figure 1. Schematic of a THz sensing platform exploiting the light-matter interaction with multiple resonances.** **(a)** Schematic of a DHR metasurface platform, which enables the static fingerprint sensing of trace pefloxacin and dynamic reaction tracking of the oxidation of vitamin C in ambient air. **(b)** Photograph of the fabricated DHR metasurface. **(c)** Microscope image of unit cells. The geometric cell parameters were fixed with a periodicity of $P_x = P_y = 250$ μm. The radii of the outer and inner rings were $r_1 = 36$ μm and $r_2 = 20$ μm, respectively. The center-to-center distance between the left and right rings was $k = 125$ μm, and the left and right rings had the same rotation angle but opposite rotation directions.

## 2.2 Multiple resonances

To understand the onset of multiple resonances, we examine the band dispersion of the metasurface (See Figure S1 in the Supporting Information), where two target resonant mode at the Γ point is found at 0.78 and 0.99 THz. The spectral response of the proposed DHR metasurface is further characterized through finite-element numerical simulations, with the corresponding field distributions presented in **Figures 2**a and 2b. Floquet periodic conditions are set in the *x*- and *y*-directions, while the open boundary condition is taken in the *z*-direction. Here, we focus on horizontal polarization (*x*-polarization) incident parallel to the opening direction ($\varphi = 0°$). Figure 2a illustrates the surface current distribution and field enhancement of the unit cell at the rotation angle of $\varphi = 30°$, where two distinct resonant modes are observed. The first mode, referred to as BIC I, exhibits a circulating current pattern, while the second mode, BIC II, features a current distribution that diverges symmetrically from the region opposite the aperture toward both tips of the resonator. These two resonances give rise to pronounced field enhancement concentrated at the ring gaps, as shown in the two figures below of Figure 2a, highlighting their strong light–matter interaction capabilities. In addition to the two primary resonances, the metasurface also supports a third high-Q quasi-BIC resonance at 1.1 THz, which is not discussed here.

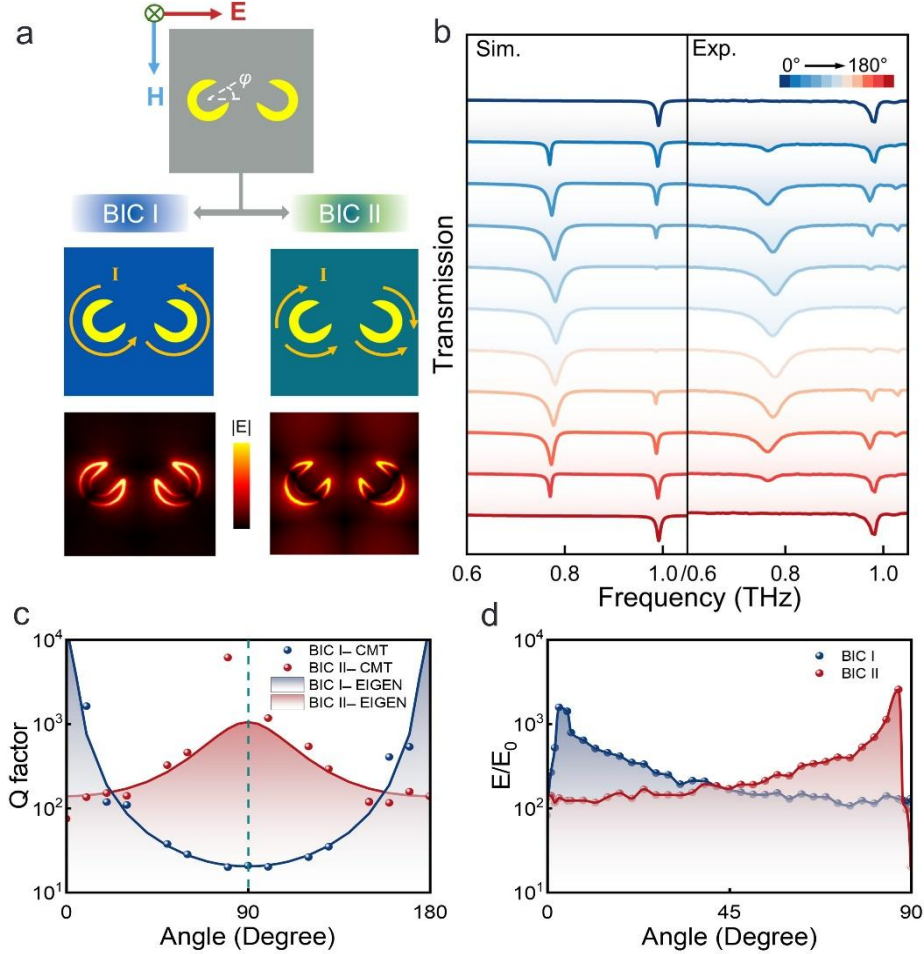

**Figure 2. Evolution and performance of multiple resonances.** (a) Schematic of a metasurface supporting multiple resonances. The unit cell consists of a pair of mirror-symmetric, horseshoe-shaped metallic unit cells, with the current distributions and field-enhanced hotspot locations of the two resonances indicated. (b) Simulated and measured transmission spectrum of the metasurface, showing excellent agreement between theory and experiment. (c) Q factors of two resonances as a function of the orientation angle. Dots represent value calculated from coupled-mode theory, and solid lines are obtained from eigenmode analysis. (d) Electric field enhancement of two resonances at various orientation angles. The Maximum field enhancement was achieved when the radiative and intrinsic Q factors were balanced.

By tuning the ring radius and aperture size, the resonance frequencies of the multiple resonances can be flexibly shifted as desired. This offers a feasible way to achieve multiple resonances of desired frequencies for THz fingerprint spectra recognition on various substances. In this work, the geometric parameters were optimized to align the resonance frequencies at 0.78 and 0.99 THz with the characteristic absorption peaks of pefloxacin in the terahertz spectrum.[49] Figure 2b illustrates the simulated transmission spectra of DHR by changing the orientation angle of the gaps. When $\varphi = 0°$, a typical symmetry-protected BIC mode appears at 0.78 THz. As the orientation angle increases, BIC I evolves into a quasi-BIC due to symmetry breaking (mirror-$y$) and ultimately transitions into a low-Q resonance at $\varphi = 90°$. Due to the change in structural symmetry, the spectral response exhibits an evolution on the contrary when the rotation angle varies from 90° to 180°. Interestingly, the evolution behavior of BIC II at 0.99 THz presents an opposite trend with BIC I, where the BIC mode occurs at $\varphi = 90°$ and disappears at $\varphi = 0°$. Compared with the existing design, our metasurface can maintain the stable resonance, with only a minor frequency shift of 0.02 THz as observed across varying orientation angles. For more detailed experimental spectral data and Coupled Mode Theory fitting curves, refer to Figure S2 in the Supporting Information.

To further quantify the evolutionary characteristics of dual resonances, we calculated their Q factors using both coupled mode theory (dots) and eigenmode calculation (line), which is shown in Figure 2c. Here, the Q-factor is defined as $Q = f_R/\text{FWHM}$, where $f_R$ represents the center frequency of the resonance, and FWHM is the full width at half-maximum of the resonant peak. The fitting results from coupled mode theory exhibit a good agreement with the simulated results of eigenmode calculation. As the rotation angle increases from 0° to 90°, the radiative Q factor of BIC I exhibits a typical angular dependence of symmetry-protected BICs, decreasing from a peak value of $1.45 \times 10^4$ to approximately 20. As the angle further increases from 90° to 180°, the Q factor symmetrically recovers to $1.45 \times 10^4$. In contrast, BIC II exhibits an opposite angular dependence, although the change in Q factor is less pronounced. As shown in Fig. 2d, the field enhancement exhibits a clear dependence on the Q factor. According to temporal coupled-mode theory, the maximum field enhancement occurs not at the highest radiative Q (BIC point) but under the critical coupling condition where radiative and intrinsic losses are balanced.[41, 50] Instead, the strongest enhancement of our design occurs at 5° for BIC I and 85° for BIC II, where an optimal balance between the radiative and intrinsic Q factors is achieved. These angle-dependent field distribution features are visualized in detail in Figure S3, which presents the electric field mode characteristics of dual resonances at specific angles, directly confirming the spatial evolution law of the field enhancement regions underlying the aforementioned enhancement behavior.

To experimentally demonstrate the transmission signature of multiple resonances, we fabricated the metasurfaces using a commercial COC layer through standard photolithography techniques (see Methods for fabrication details). The optical image and partially enlarged image of the fabrication sample can be seen in Figs. 1b and 1c. A home-built THz time-domain spectroscopy (TDS) system was employed to characterize the metasurface performance (see Methods for experimental setup). Figure 2b shows the measured transmission spectra of the fabricated metasurfaces, exhibiting excellent agreement with the simulated results. The resonance frequencies and Q factors of dual resonances both follow the characteristic dependence of symmetry-protected BICs on the orientation angle.

To analyze the multipole nature of multiple resonances, we performed the Cartesian multipole decomposition. **Figure 3**a displays the coupling mechanism of BIC I, where channel I (red arrow) represents the bright state |1⟩ (electric dipole, ED), and channel II (blue arrow) corresponds to the dark state |2⟩ (out-of-plane magnetic dipole, MD). This broken structural symmetry facilitates the interaction between the pump wave and the bright state, enabling the external excitation to couple into the dark state through the bright state. When the structure rotation angle is set as $\varphi = 0°$, MD does not couple with the incident light due to symmetry constraints. However, as $\varphi$ increases, the symmetry is broken, allowing the ED channel to be directly excited by the external plane wave. The energy accumulated in the dark state |2⟩ couples to the bright state |1⟩ through the overlap of their mode fields, enabling partial energy transfer. Meanwhile, the increased overlap between the external pump and MD enhances the coupling strength, leading to a broader resonance linewidth and a reduced Q factor of the resulting quasi-BIC. Furthermore, BIC II follows the same coupling mechanism, except that the ideal BIC condition occurs at a rotation angle of $\varphi = 90°$.

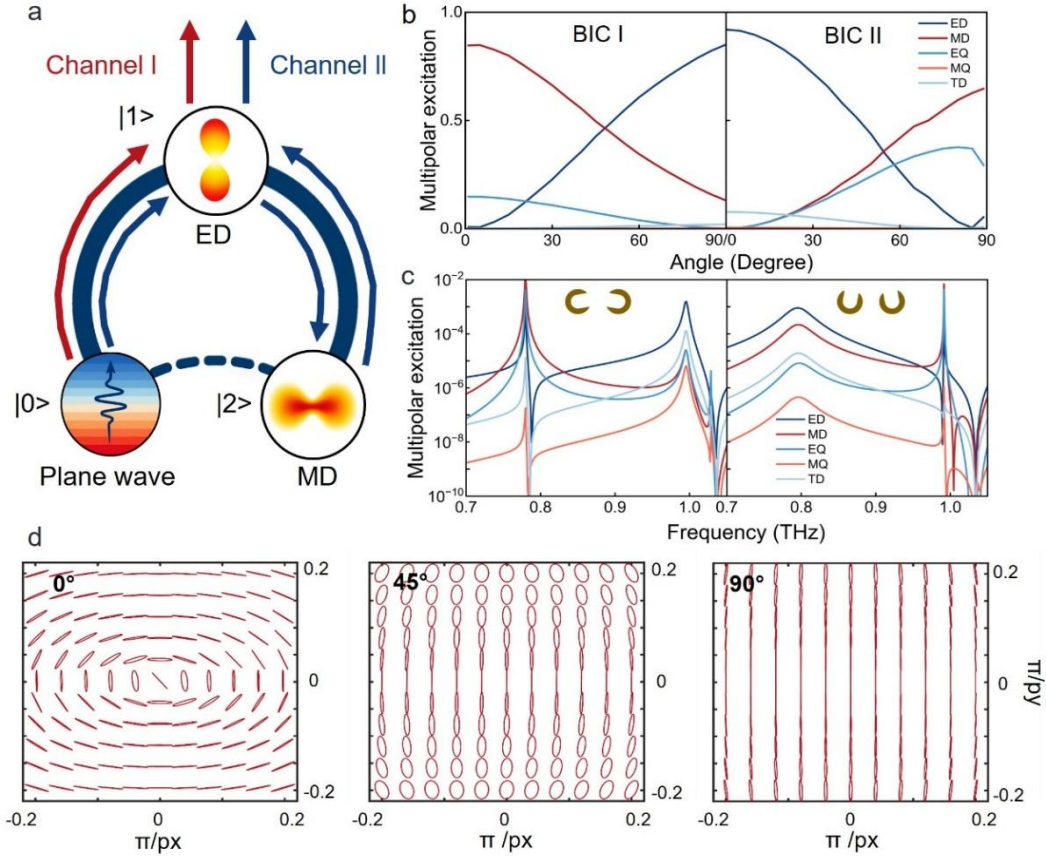

**Figure 3. Multipole decomposition and far-field polarization distributions of multiple resonances.**
**(a)** Schematic of the coupling mechanism. State |0⟩ denotes x-polarized excitation, |1⟩ corresponds to the electric-dipole bright state, and |2⟩ to the magnetic-dipole dark state. Solid arrows indicate energy flow, colored lines denote two distinct radiation channels, and dashed arrows represent forbidden couplings.
**(b)** Multipole decomposition at the resonance frequency for various incidence angles $\varphi$. **(c)** Multipole decomposition across different frequencies at $\varphi = 10°$ and 80°. **(d)** Far-field polarization distributions under oblique incidence angles of 0°, 45°, and 90°.

Figure 3b gives the multipolar decomposition of multiple resonances with the orientation angle increasing from 0° to 90°. Taking the first one as an example, at $\varphi = 0°$, a symmetry-protected MD mode emerges at the Γ point of the metasurface, forming an ideal BIC. As $\varphi$ increases, symmetry breaking introduces a leaky channel via ED excitation, leading to a gradual decrease in the Q factor of BIC I. In contrast, for the second, an opposite trend between the ED and MD contributions is observed as the orientation angle varies from 0° to 90°, with the electric dipole (ED) mode serving as the dominant resonance. To illustrate the multipole characteristics of different resonances at different angles, Figure 3c displays the multipole response of the metasurface at rotation angles of 10° and 80°, where both the multipolar composition and the linewidth evolution of the dual resonances can be simultaneously observed. Specifically, at small orientation angles, the first resonance exhibits a narrower linewidth, whereas the second shows a broader linewidth. Conversely, at larger angles, the linewidth of the first resonance broadens, while the second becomes narrower. A more comprehensive set of multipolar radiation patterns across additional angles (5° to 85°) is provided in Figure S4, which further captures the continuous angular evolution of spectral characteristics for all five multipolar components (ED, MD, EQ, MQ, TD) and reinforces the angle-dependent trends in linewidth described above.

Furthermore, the far-field polarization map of the first resonance is shown in Figure 3d, where the dimensionless wave vector components $\pi/p_x$ and $\pi/p_y$ are employed as the $x$ and $y$ axes, respectively. At each momentum point, the major axis direction and flattening of the polarization ellipse reflect the principal direction and ellipticity, respectively. As the rotation angle $\varphi$ increases from 0° to 90°, BIC I undergoes a continuous evolution in far-field polarization—from a perfect vortex to a twisted vortex, and finally to nearly parallel stripes. At $\varphi = 0°$, the polarization ellipse completes a $2\pi$ rotation around the singularity point, forming a closed vortex with a topological charge $|q| = 1$. The polarization direction is undefined at the vortex core, indicating a complete decoupling of the BIC mode from the radiation continuum and resulting in a theoretically infinite Q factor. As $\varphi$ increases to 45°, the vertical mirror symmetry is broken, and the vortex ring is stretched into an ellipse and slightly shifted, revealing anisotropy in the coupling efficiencies between ED and MD. At $\varphi = 90°$, the structural symmetry is completely broken, and the major axes of the polarization ellipses align, forming nearly parallel stripes. This evolution clearly illustrates how symmetry breaking governs the interplay between ED and MD coupling strengths and modulates the excitation of radiative leakage channels.

## 2.3 THz fingerprint detection of Pefloxacin

To leverage the strong terahertz–matter interaction enabled by multiple resonances, we implement the sensing experiment of the proposed metasurface to improve molecular fingerprint detection at terahertz frequencies. The metasurface is capable of supporting multiple BIC resonances, but this work specifically focuses on dual-channel detection optimized to align with the two primary fingerprint peaks of pefloxacin at 0.78 and 0.99 THz. By deliberately matching multiple high-Q metasurface resonances to the intrinsic fingerprint absorption frequencies of pefloxacin, strong localized field enhancement amplifies the weak molecular absorption. More details can be seen from the supporting information. **Figure 4**a provides an overview of the experimental procedure for molecular fingerprint detection, in which the metasurface was mounted on a perforated backplate. The measured signal is measured from the back of the sensors on the THz-TDS, and the environmental humidity was controlled below 3% throughout the experiment to eliminate environmental interference. Pefloxacin (PEF), a molecule with two distinct absorption features in the terahertz frequency range, was used as the test sample. Prior to sample preparation, the transmission spectrum of the bare metasurface was measured as a reference under the same conditions described in Fig. 2b. Subsequently, 20 μL of PEF solution at various concentrations was deposited onto the metasurface (see Methods for details). After drying, the PEF molecules are uniformly distributed across both the unit cells and the surrounding substrate areas.

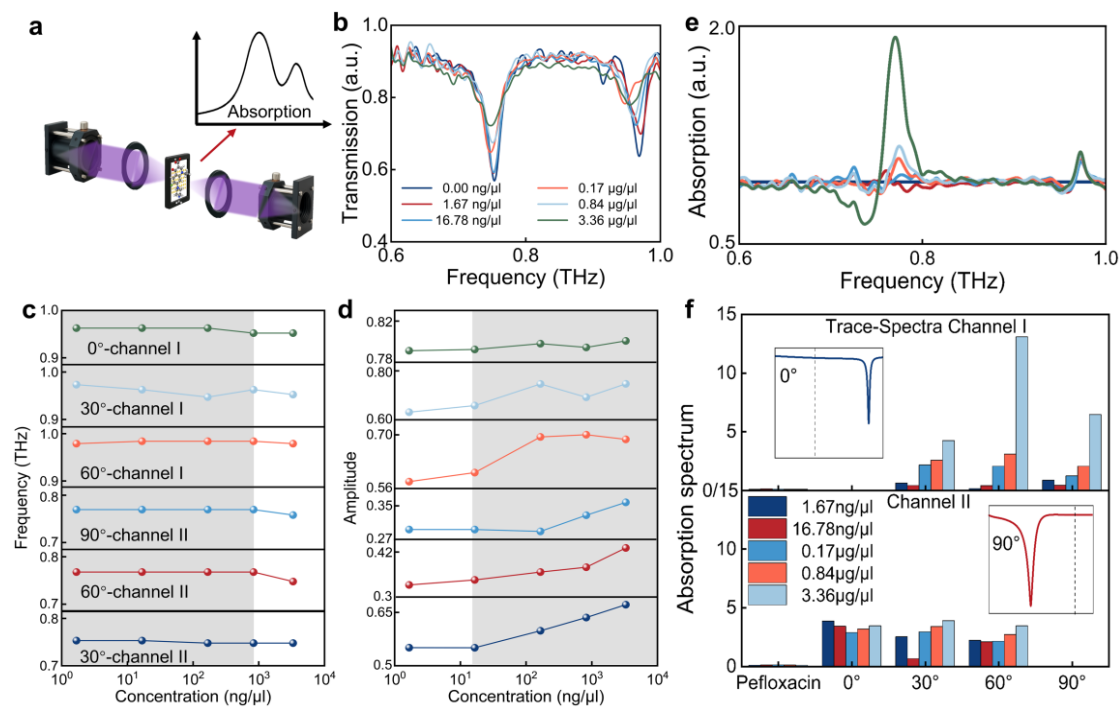

**Figure 4. THz fingerprint spectral sensing of multiple resonances. (a)** Schematic of the experimental setup used for pefloxacin fingerprint-spectrum measurements. **(b)** Measured transmission spectra as functions of pefloxacin concentration. **(c)** Resonant-frequency shifts of six sensing channels across four metasurfaces at changing concentration. Two resonances per metasurface are treated as two independent channels; frequency shifts are negligible at low concentrations but become pronounced at higher concentrations. **(d)** Amplitude changes of six sensing channels versus concentration, showing little variation at low concentrations and significant change at higher concentrations. **(e)** Extracted fingerprint spectra of the pefloxacin-containing metasurfaces. All curves were measured on metasurfaces with the orientation angle of 60°, with sample concentrations identical to those in (b). **(f)** Absorption enhancement of the two sensing channels for the metasurface, relative to the pefloxacin COC layer, at different incidence angles. Channel 1 is absent at 0°, and Channel 2 is absent at 90°

In our experiments, four metasurfaces with orientation angles of 0°, 30°, 60°, and 90° were used to examine five different concentrations of PEF (1.67ng/μl, 16.78ng/μl, 0.17ug/μl, 0.84ug/μl and 3.36ug/μl). Figure 4b demonstrates the transmission spectra of pefloxacin at different concentrations with $\varphi = 60°$. As the sample concentration increases, the spectral response exhibits two notable features: an increase in absorption amplitude and a minor redshift of the resonance frequency, both of which are consistent with theoretical expectations. It is worth noting that increasing pefloxacin concentration introduces additional intrinsic absorption loss that perturbs the loss balance of the quasi-BIC metasurface, leading to linewidth broadening and contrast reduction. To quantitatively analyze the dependence of frequency shift and amplitude of measured results on the concentrations of PEF, we extracted the corresponding data of four metasurfaces. Figures 4c and 4d illustrate the detection limit of fingerprint retrieval by identifying the lowest concentration at which the characteristic features remain discernible. Here, BIC I and BIC II are defined as Channel I and Channel II, respectively. At lower concentrations (marked by the grey area), all the channels with different orientation angles show a negligible frequency shift. Taking a frequency shift of 0.01 THz as the threshold (determined by the frequency resolution of the THz-TDS system), all channels start to exhibit an obvious red shift when the concentration exceeds $10^3$ ng/μL. While for the amplitude shown in Figure 4d, there is no significant change that occurs at the two channels when the concentrations are below 10 ng/μL. At higher concentrations (highlighted by the grey-shaded region), a clear upward trend is observed, with the amplitude increasing by up to 0.2. Repeated measurements confirm that the amplitude deviation is within ±0.02, ensuring the reliability of the sensing results. The accumulation of PEF at high concentrations leads to pronounced frequency shifts and amplitude enhancements across the channels.

Beyond the conventional study focus on frequency and amplitude responses, we further performed precise extraction of the absorption spectra corresponding to five different concentrations of PEF using multiple resonances, as shown in Figure 4e. The absorption-related spectrum in Fig. 4e is obtained by frequency-dependent normalization of the experimental data, where the transmission of the analyte-loaded metasurface is divided by that of the blank reference to isolate the relative changes induced by pefloxacin molecules. We found that two absorption peaks around 0.78 and 0.99 THz arise from the spectrum of all concentrations, which exactly match the fingerprint absorption features of PEF. Interestingly, the absorption peaks show a monotonic increase in amplitude with concentration. Due to the low concentration of the analyte and the relatively thin molecular layer, no resonance splitting is observed. For a more comprehensive visualization of the spectral responses across different metasurface configurations, Fig. S5 presents the fingerprint spectra under four orientation angles and the same five PEF concentrations (including a reference spectrum without the metasurface), clearly illustrating the angle-dependent variation in transmission spectra that underpins the absorption characteristics discussed above.

For the quantitative analysis of absorption peaks, Figure 4f shows the extracted measured absorption data of channels I and II, respectively. The quantification of absorption enhancement is established by referencing the amplitude variation ($\Delta a/a$) in the transmittance spectrum of the PEF-coated COC substrate ($a$ denotes the transmittance amplitude of the pure COC substrate, and $\Delta a$ represents that of the PEF-coated COC substrate). Specifically, the absorption enhancement is defined as the ratio of the change observed in the metasurface configuration, thus computed by dividing the response of the proposed metasurface by the reference: $(\Delta b/b)/(\Delta a/a)$. Notably, the metasurface demonstrates substantial absorption enhancement across the full absorption peak of PEF, with a maximum enhancement factor reaching 13.06. Moreover, it exhibits a distinct Q-factor-dependent behavior: Channel I is inactive at a rotation angle of 0°, while Channel II becomes inactive at 90°. The integration of multiple resonances within a single-pixel metasurface enables successful fingerprint retrieval of PEF, thereby demonstrating the unique advantages of multiple resonance systems in enhancing light–matter interactions. This approach not only overcomes the limitations of single-resonance-based sensing but also expands the dimensionality of spectral recognition.

**2.4 Real-time monitoring of the vitamin C oxidation**

To extend the applicability of the proposed designs, we also investigate the dynamic biochemical reaction tracking capabilities of the metasurface-based sensing platform. Here, we employed the non-enzymatic oxidation process of vitamin C (VC) as a representative biochemical reaction and analyzed its impact on terahertz spectral response. As shown in **Figure 5**a, the experimental system utilizes a transmission THz spectroscopy system integrated with temperature control components, ensuring that the VC oxidation reaction proceeds under room-temperature conditions. The preparation of VC deposited onto the metasurface could be seen in the methods. Upon exposure to air, VC molecules undergo stepwise hydrogen atom abstraction from the enediol group by molecular oxygen, leading to the formation of a semidehydroascorbic acid radical intermediate, which subsequently converts into dehydroascorbic acid (DHAA). The reaction follows second-order kinetics with a characteristic timescale of several hours. As the spectral changes became negligible after approximately 1 hour, all subsequent measurements were conducted within this time frame. This well-characterized non-enzymatic oxidation pathway serves as an ideal model for studying dynamic molecular interactions in the terahertz regime. Notably, measured resonance frequency remained at $0.62 \pm 0.01$ THz throughout oxidation (marked by the dashed line in Figure 5a). This stability is attributed to the insensitivity of the metasurface unit cell's electromagnetic resonance to the weak dielectric perturbations induced by the reaction, thereby providing a robust reference point for subsequent quantitative analysis based on amplitude variations.

Figure 5b presents the time-resolved spectral evolution obtained from the metasurface with an orientation angle of $\varphi = 90°$, under a VC concentration of 25 μg/μL. Given the stable resonance frequency, equidistant frequency-axis translation with 0.01 THz intervals was applied to the original spectra, effectively separating overlapping multi-peak signals. This analysis revealed a monotonic amplitude decay, which is directly associated with the reduction in molecular polarizability induced by VC oxidation. Detailed dynamic spectral evolution under additional parameter combinations, including original, restored, and time-dependent spectra is provided in Figure S6, further illustrating the angle and concentration-dependent differences in transmission amplitude changes during VC oxidation. In contrast, Figure 5c shows the response curve of $\varphi = 60°$ and VC concentration of 10 μg/μL. A similar amplitude decay was observed, confirming the reusability of the metasurface. Post-reaction transmission spectra showed that, after cleaning, the transmission peak intensity fully recovered to its initial value, indicating that the reaction caused no irreversible damage to the electromagnetic properties of the metasurface. This reusability is further validated by morphological evidence in Figure S7, which captures the metasurface's structural integrity before interaction, the traces of VC adsorption, and its recovery after cleaning

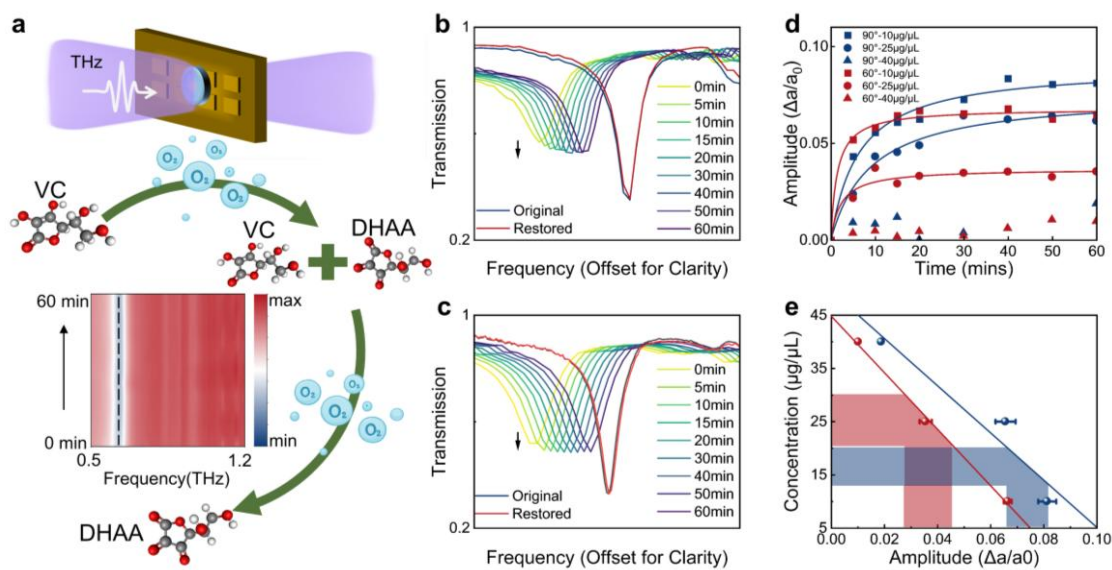

**Figure 5. THz dynamic biochemical reaction monitoring using multiple resonances.** (**a**) Schematic of the dynamic monitoring setup and time–frequency spectrogram of the vitamin C oxidation process. (**b**) Time-resolved transmission spectra for 25 μg/μL vitamin C on a metasurface with the orientation angle of 90°. As illustrated in (a), the resonance frequency remains essentially fixed; the spectra have been offset along the frequency axis to emphasize amplitude variations. (**c**) Time-resolved transmission spectra for 10 μg/μL vitamin C on a metasurface with the orientation angle of 60°. (**d**) Extracted time-dependent amplitude changes at the resonance frequency for six experimental runs; curve fitting was performed on four of the datasets. (**e**) Estimation of the initial vitamin C concentration based on the final amplitude change ratios.

To quantitatively track the amplitude evolution, we extracted the normalized amplitude change rate ($\Delta a/a_0$, where $\Delta a$ and $a_0$ denote the amplitude variation and initial amplitude, respectively) at each time point, enabling the construction of kinetic curves that reflect the progression of the oxidation process, which is shown in Figure 5d. Over time, the amplitude under all tested conditions rises initially before reaching a steady state. It is obvious that the amplitude response associated with $\varphi = 90°$ surpasses that of $\varphi = 60°$ at an identical concentration. Furthermore, for a given orientation, higher concentrations correspond to elevated steady-state amplitude values. The observed saturation behavior of vitamin C oxidation reflects the limited sensing volume and site occupancy at the metasurface hotspots, where the field-enhancement regions act as virtual enzymes and are quantitatively monitored via the normalized amplitude change rate ($\Delta a/a_0$). The fitting results for curves at different concentrations, shown as solid lines in Figure 5d, were obtained using the Michaelis–Menten model (details could be seen from supporting information), a commonly used kinetic framework for describing enzyme-substrate interactions.[51] The coefficient of determination $R^2 > 0.92$ (0.9837, 0.97644, 0.98862, 0.92662) indicates that the model is highly consistent with the experimental data, verifying the reliability of the obtained kinetic parameters. More detailed kinetic parameters, including $V_{max}$ and $K_m$ values, are presented in Table S1. The fitting results yield two key conclusions: i) The oxygen concentration in the reaction system is significantly higher than that of VC, thereby satisfying the quasi–steady-state approximation condition commonly used in Michaelis–Menten kinetics.[52] ii) The field-enhancement hotspots generated by the metasurface resemble enzyme–substrate interactions in their ability to capture, adsorb, and facilitate the reaction of VC molecules.[53]

Furthermore, as shown in Fig. 5e, based on the characteristic amplitude change ($\Delta a/a$) following complete oxidation and the established concentration–response calibration curve, label-free quantitative estimation of the initial VC concentration can be achieved within a defined concentration range. Specifically, by mapping the measured $\Delta a/a$ value to the pre-constructed calibration curve, which correlates $\Delta a/a$ with known VC concentrations, the corresponding initial concentration of the analyte can be directly retrieved. Compared with conventional techniques that focus solely on static substance sensing and identification, this platform leverages the dynamic terahertz response to continuously track the evolution of dielectric properties during biochemical reactions. This capability offers distinct advantages for dynamic applications such as food freshness monitoring and drug metabolism analysis.

## 3. Conclusion

We have introduced a fundamentally different approach to molecular fingerprinting that departs from conventional spatially multiplexed metasurfaces. By integrating multiple quasi-BIC resonances within a single-pixel metasurface, we eliminate the need for large-scale pixelated arrays and laborious spatial scans. We have developed a novel platform for metasurface sensing that explores multiple resonances produced by quasi-bound states in the continuum. Our platform facilitates simultaneous excitation of two resonances, in particular at 0.78 and 0.99 THz, offering enhanced spectral flexibility and substantial field confinement. We have demonstrated the efficiency of this multiple resonance platform for advanced sensing performance on two representative examples. For static fingerprint retrieval, the metasurface enables label-free identification of trace-level pefloxacin via distinct multispectral absorption fingerprints, achieving a maximum signal enhancement factor of 13.06. For dynamic biochemical reaction tracking, this metasurface platform captures real-time oxidation kinetics of vitamin C by monitoring resonance amplitude variations driven by strong near-field enhancement. The extracted kinetic profiles exhibit excellent agreement with nonlinear models ($R^2 > 0.92$), and the structure demonstrates high reusability with complete spectral recovery across multiple cycles. By combining static fingerprint retrieval and dynamic biochemical reaction tracking, our multiresonant platform surpasses the limitations of conventional single-resonance sensors. Thus, our versatile sensing platform offers scalable and high-performance solutions for biochemical sensing, with promising applications in areas such as drug residue detection and food quality monitoring.

## 4. Experimental Section

*Numerical approach and tools:* The eigenfrequencies and Q factors of multiple resonances are simulated via eigenmode analysis in COMSOL Multiphysics, incorporating periodic boundary conditions. Additionally, CST Microwave Studio (utilizing a frequency-domain solver with unit cell conditions) is used to characterize the transmission spectra and field distributions of the dual C-shaped ring structure.

*Metasurface Fabrication:* The dual C-shaped ring structure employs a commercial 50-μm-thick cyclic olefin copolymer (COC) film as the substrate, which exhibits low loss tangent characteristics. Firstly, the COC layer was attached to a 500-μm-thick silicon wafer to serve as a rigid support, preventing deformation during the fabrication process. Standard photolithography was used for patterning the metasurface: PR4000 photoresist was spin-coated onto the COC film to form a resist layer approximately 3 μm thick. After a 2-minute pre-baking, the sample was exposed to ultraviolet light through a photomask with the dual C-shaped ring pattern, followed by a 90-second development to form the photoresist pattern. A 200-nm-thick copper layer was deposited on the COC surface under vacuum using magnetron sputtering. Subsequently, the sample was immersed in an acetone solution for 10 minutes for the lift-off process, selectively removing the copper layer on the photoresist to form the desired dual C-shaped ring structure. After removing acetone and cleaning the surface with isopropyl alcohol and deionized water, the COC film was peeled off the silicon wafer.

*Experimental characterization:* In the experiment, the samples are characterized using a photoconductive antenna-based THz time-domain spectroscopy (THz-TDS) system under normal incidence. An x-direction polarizer is placed in front of the sample to ensure high polarization directionality of the incident wave. The THz-TDS system is configured with a 190-ps delay line (0.02-ps precision), three averaging scans, and dry air purging to eliminate absorption noise from water vapor.

*Solution Preparation:* <u>Pefloxacin solutions.</u> Pefloxacin (cat. no. P815514, 99% purity, Macklin Biochemical Co., Ltd., Shanghai) was used for solution preparation. The specific steps were as follows, using an analytical balance with a precision of 1 mg, 1 mg, 10 mg, 100 mg, 500 mg, and 2000 mg of pefloxacin powder were accurately weighed, respectively. Considering the 99% purity, the actual masses of the active ingredient were 0.99 mg, 9.9 mg, 99 mg, 495 mg, and 1980 mg, which were added to 596 mL of deionized water separately. The mixture was stirred at room temperature (25±1 °C) to ensure complete dissolution of pefloxacin, resulting in five solutions with different concentrations. Immediately after preparation, 20 μL of each pefloxacin solution was precisely pipetted (pipette range: 20–200 μL, precision: ±0.1 μL) and added dropwise to the central area of the metasurface. During pipetting, the tip of the pipette was kept 1 cm away from the metasurface to avoid liquid splashing. The metasurface with the dropped solution was placed horizontally on an 80 °C hotplate for 8–10 minutes to allow complete evaporation of the solvent. After drying, the metasurface was immediately transferred to the sample stage of the aforementioned THz-TDS system for characterization.

Vitamin C solutions. L-ascorbic acid (cat. no. A800296, metal impurity content ≤0.01%, Macklin Biochemical Co., Ltd., Shanghai) was used for solution preparation. The main difference in preparation was that after pipetting the solution onto the central area of the metasurface, the metasurface with the dropped solution was placed in a vacuum oven and dried at 40 °C (temperature fluctuation: ±0.5 °C) for 30 minutes, with stable pressure maintained inside the oven during drying. Subsequently, it was transferred to the THz-TDS system for testing, with dry air purging throughout the process.

**Supporting Information**

Supporting Information is available from the Wiley Online Library or from the author.

**Acknowledgments**

The authors thank Dr. Mingkai Liu for helpful discussions. This work was supported by the National Natural Science Foundation of China (Grant No. 62205380) and Natural Science Foundation of Hunan Province (Grant No. 2024JJ6529). Y.K. acknowledges support from the Australian Research Council (Grant No DP210101292) International Technology Center Indo-Pacific (ITC IPAC) via Army Research Office (contract FA520923C0023). Access to facilities used in this work is made possible through the Facility for Micro/Nano Fabrication and Device Manufacture, Central South University.

**Conflict of Interests**

All other authors declare they have no competing interests.

**Author contributions:**
Conceptualization: Q.Y., Y.D., Y.Z., J.H., Y.K.
Methodology: Q.Y., Y. D., Y.Z., L.X.
Investigation: Q.Y., J.H., F.L., Y.Z., Q.X.
Visualization: Q.Y. Y.D.
Supervision: Q. Y., J. Y., J.H., I. S., Y.K.
Writing—original draft: Q.Y., Y.D.
Writing—review & editing: D.W., C.L., J.Y., J. H., I.S., Y.K.

**Data availability statement**

The data that support the findings of this study are available from the corresponding author upon reasonable request